\documentclass[12pt,epsf]{article}
\usepackage{amsmath,amssymb,graphicx}
\begin{document}
\def\beq{\begin{eqnarray}}
\def\eeq{\end{eqnarray}}
\newcommand{\gsim}{ \mathop{}_{\textstyle \sim}^{\textstyle >} }
\newcommand{\lsim}{ \mathop{}_{\textstyle \sim}^{\textstyle <} }
\newcommand{\vev}[1]{ \left\langle {#1} \right\rangle }

\baselineskip 0.7cm

\begin{titlepage}

\begin{flushright}
UT-09-14
\\
IPMU-09-0060
\end{flushright}

\vskip 1.35cm
\begin{center}
{\large \bf
Inverse Problem of Cosmic-Ray Electron/Positron\\
 from Dark Matter
}
\vskip 1.2cm
Koichi Hamaguchi$^{1,2}$, Kouhei Nakaji$^1$ and Eita Nakamura$^1$
\vskip 0.4cm

{\it {}$^1$ Department of Physics, University of Tokyo, Tokyo 113-0033, Japan\\
{}$^2$ Institute for the Physics and Mathematics of the Universe, 
University of Tokyo, Chiba 277-8568, Japan
}

\vskip 1.5cm

\abstract{
We discuss the possibility of solving the inverse problem of the cosmic-ray electron/positron from decaying/annihilating dark matter,
and show simple analytic formulae to reconstruct the source spectrum of the electron/positron from the observed flux.
We also illustrate our approach by applying the obtained formula to the just released Fermi data as well as the new HESS data.
}
\end{center}
\end{titlepage}

\setcounter{page}{2}

\section{Introduction}
The existence of Dark Matter (DM) is by now well established~\cite{DM}, yet its identity is a complete mystery; it has no explanation
in the framework of the Standard Model of particle physics.

Recently, several exciting data have been reported on cosmic-ray electrons and positrons, which may be indirect signatures of
decaying/annihilating DM in the present universe.
The PAMELA collaboration reported that the ratio of positron and electron fluxes increases
at energies of $\sim$ 10--100 GeV~\cite{Adriani:2008zr}, which shows an excess above
the expectations from the secondly positron sources.
In addition, the ATIC collaboration
reported an excess of the total electron + positron flux at energies between 300 and 800 GeV~\cite{:2008zzr}. (cf. the PPB-BETS
observation~\cite{Torii:2008xu}.)

More recently, the Fermi-LAT collaboration has just released precise, high-statistics data on cosmic-ray electron + positron spectrum
from 20 GeV to 1 TeV~\cite{Fermi}, and the HESS collaboration has also reported new data~\cite{HESS}. Although these two new data
sets show no evidence of the peak reported by ATIC, their spectra still indicate an excess above conventional models for the background
spectrum~\cite{Fermi}, and the decaying/annihilating DM remains an interesting possibility~\cite{post-Fermi}.

In the previous studies of DM interpretations of the cosmic-ray electrons/positrons, 
the analyses have been done by (i) assuming a certain source spectrum from
annihilating/decaying DM, (ii) solving the propagation and (iii) comparing the predicted electron and positron fluxes with
the observation. In this letter, we discuss the possibility of solving its inverse problem, namely, try to reconstruct the
source spectrum from the observational data. For that purpose, we adopt an analytical approach, since the inverse problem
would be very challenging in the numerical approach.

We show that the inverse problem can indeed be solved analytically under certain assumptions and approximations,
and provide analytic formulae to reconstruct the source spectrum of the electron/positron from the observed flux. 
As an illustration, we apply the obtained formulae to the electron and positron flux above $\sim 100$ GeV for the just released Fermi
data~\cite{Fermi}, together with the new HESS data~\cite{HESS}.
It is found that the reconstructed spectrum for high energy range is almost independent of the propagation models
and whether the DM is decaying or annihilating.
We also estimate the errors of the reconstructed source spectrum.
The obtained result implies that electrons/positrons at the source have a broad spectrum ranging from ${\cal O}(1)$ TeV to ${\cal O}(100)$ GeV,
with a broad but clear peak around $E\sim 400$ GeV and another peak at $E\sim 1$ TeV.

\section{Inverse problem of cosmic-ray electron and\\ positron from dark matter}
Let us first summarize the procedure to calculate the $e^\pm$ fluxes. The electron/positron number density
per unit kinetic energy $f_e(E,\vec{r},t)$ evolves as~\cite{Strong:2007nh}
\begin{eqnarray}
\frac{\partial}{\partial t}f_e(E,\vec{r},t) 
&=& 
\nabla\cdot [K(E,\vec{r})\nabla f_e(E,\vec{r},t)] 
+ \frac{\partial}{\partial E}[b(E,\vec{r}) f_e(E,\vec{r},t)]
+Q(E,\vec{r},t)\,,
\end{eqnarray}
where $K(E,\vec{r})$ is the diffusion coefficient, $b(E,\vec{r})$ is the rate of energy loss, and
$Q(E,\vec{r},t)$ is the source term of the electrons/positrons.
The effects of convection, reacceleration, and the annihilation in the Galactic disk are neglected.
We only consider the electrons and positrons from dark matter decay/annihilation, which we assume time-independent and spherical;
\begin{eqnarray}
Q(E,\vec{r}) = q(|\vec{r}|) \frac{dN_e(E)}{dE}\,,
\end{eqnarray}
where $dN_e/dE$ is the energy spectrum of the electron and positron from one DM decay/annihilation, 
and $q(|\vec{r}|)$ is given by
\begin{eqnarray}
q(|\vec{r}|) &=& \frac{1}{m_X\tau_X}\rho(|\vec{r}|)\quad{\rm for~decaying~DM},
\\
q(|\vec{r}|) &=& \frac{\vev{\sigma v}}{2m_X^2}\rho(|\vec{r}|)^2\quad{\rm for~annihilating~DM},
\end{eqnarray}
where $m_X$ and $\tau_X$ are the mass and the lifetime of the DM, and $\vev{\sigma v}$ is the average DM annihilation cross section.
We adopt a stationary two-zone diffusion model with cylindrical boundary conditions, 
with half-height $L$ and a radius $R$, and spatially constant diffusion coefficient
$K(E)$ and the energy loss rate $b(E)$ throughout the diffusion zone. The diffusion equation is then
\begin{eqnarray}
K(E)\nabla^2 f_e(E,\vec{r}) + \frac{\partial}{\partial E}[b(E)f_e(E,\vec{r})] + q(|\vec{r}|) \frac{dN_e(E)}{dE} = 0\,,
\label{eq:diffusion}
\end{eqnarray}
with the boundary condition $f_e(E,\vec{r})=0$ for $r=\sqrt{x^2+y^2}=R$, $-L\le z\le L$ and $0\le r \le R$, $z=\pm L$.
As shown in Appendix A, its solution at the Solar System, $r=r_\odot\simeq 8.5$ kpc, $z=0$ is given by
\begin{eqnarray}
f_e(E) = f_e(E,\vec{r}_\odot)
=\frac{1}{b(E)}\int^{E_{\rm max}}_EdE'
\frac{dN_e(E')}{dE'}
g\left(L(E')-L(E)\right),
\label{eq:dN/dE}
\end{eqnarray}
where
\begin{eqnarray}
g(x)&=&\sum_{n,m=1}^\infty
J_0 \left(j_n \frac{r_\odot}{R}\right)
\sin\left(\frac{m\pi}{2}\right)
q_{nm}e^{-d_{nm}x},
\label{eq:gx}
\\
q_{nm}&=&
\frac{2}{J_1^2(j_n) \pi}
\int^1_0 d\hat{r}\, \hat{r}
\int^\pi_{-\pi} d\hat{z}\,
J_0(j_n \hat{r})\sin\left(\frac{m}{2}(\pi-\hat{z})\right)
q\left(\sqrt{(R\hat{r})^2 + \left(\frac{L\hat{z}}{\pi}\right)^2}\right),
\label{eq:qnm}
\\
d_{nm} &=& \frac{j_n^2}{R^2}+\frac{m^2 \pi^2}{4L^2},
\label{eq:dnm}
\\
L(E) &=& \int^EdE'\frac{K(E')}{b(E')},
\label{eq:L}
\end{eqnarray}
where $j_n$ are the successive zeros of $J_0$. The electron/positron flux is given by $\Phi_e(E) = (c/4\pi)f_e(E)$.

Our main purpose is to solve the inverse problem, namely, to reconstruct the source spectrum $dN_e(E)/dE$ for
a given $f_e(E)$. As shown in Appendix \ref{inv_sol}, this inverse problem can indeed be solved, and the solution is given by
\begin{eqnarray}
\label{solution_Fourier} 
\frac{dN_e(E)}{dE}
&=&
\frac{dL(E)}{dE}
\int_{-\infty}^{\infty}
\frac{dk}{2\pi}
e^{-ikL(E)}
\frac{\int_{-\infty}^{\infty}dw
e^{-ikw}\tilde{A}(w)}{\int_{0}^{\infty}dz
e^{-ikz}g(z)},
\end{eqnarray}
where 
\begin{eqnarray}
A(E)=f_e(E)b(E)
\end{eqnarray}
and
\begin{equation}
\tilde{A}(x=-L(E))=A(E),
\end{equation}
or alternatively,
\begin{eqnarray}
\label{solution_Laplace}
\frac{dN_e(E)}{dE}=-\frac{1}{g(0)}\left[\frac{dA(E)}{dE}+
\frac{dL(E)}{dE}\int_E^{E_{\rm max}}dE'\,\frac{dA(E')}{dE'}\Gamma\left(L(E')-L(E)\right)\right],
\end{eqnarray}
where the function $\Gamma(x)$ is determined from $g(x)$. See Appendix \ref{inv_sol} for details.

Several comments are in order. First of all, from Eq.~(\ref{solution_Laplace}) one can see that the source flux at an energy
$E_{\rm src}$ depends only on the observed flux at $E_{\rm obs}\geq E_{\rm src}$. 
This may be counterintuitive, because the solution to the diffusion equation (\ref{eq:dN/dE}) tells us that the observed flux at $E=E_{\rm obs}$
is determined by the source flux at $E_{\rm src}\geq E_{\rm obs}$. It can be understood by considering the reconstruction of
the source spectrum from the highest energy and gradually to the lower energy.

Secondly, as we will see in the explicit examples, for high energy range 
the above formulae can be approximated by the first term of Eq.~(\ref{solution_Laplace});
\begin{eqnarray}
\frac{dN_e(E)}{dE}\simeq -\frac{1}{g(0)} \frac{dA(E)}{dE}\,.
\label{eq:simple}
\end{eqnarray}
This is because at higher energy the electrons lose their energy quickly and hence only local electrons contribute. Technically, at higher
energy $g(x)$ can be approximated as $g(x)\simeq g(0)$. 
The approximated formula Eq.~(\ref{eq:simple}) is then directly obtained from Eq.~(\ref{eq:dN/dE}). Note that $g(0)$ is given by the local
DM density $\rho_\odot\simeq 0.30$~GeV/cm$^3$;
\begin{eqnarray}
g(0) &=& q(r=r_\odot, z =0)
=\left\{
\begin{array}{ll}
\displaystyle{\rho_\odot\frac{1}{m_X\tau_X}} & {\rm for~decaying~DM},
\\
&
\\
\displaystyle{\rho_\odot^2\frac{\vev{\sigma v}}{2 m_X^2}} & {\rm for~annihilating~DM}.
\end{array}
\right.
\end{eqnarray}

\section{Examples}
In this section, we show some examples. 
The energy loss rate is taken as $b(E) = E^2/E_0\tau_E$, with $E_0=1$ GeV and $\tau_E=10^{16}$ sec,
and the propagation models are parametrized by
$K(E) = K_0(E/E_0)^\delta$ which leads to
\begin{eqnarray}
L(E) = -\frac{\tau_E K_0}{1-\delta}\left(\frac{E_0}{E}\right)^{1-\delta}.
\end{eqnarray}
\begin{table}[t]
\begin{center}
\begin{tabular}{|c|cc|cc|} \hline
Models & $R$ [kpc] & $L$ [kpc] & $\delta$ & $K_0$ [kpc$^2/$Myr] \\ \hline
M2     &  20 & 1  & 0.55    & 0.00595             \\ 
MED    &  20 & 4  & 0.70    & 0.0112             \\
M1     &  20 & 15 & 0.46    & 0.0765             \\ \hline
\end{tabular}
\caption{The diffusion model parameters consistent with the observed B/C ratio~\cite{Delahaye:2007fr}.}
\end{center}
\end{table}
We consider the three benchmark models from Ref.~\cite{Delahaye:2007fr}, M2, MED, and M1, which are summarized in Table~1.
The parameters $K_0$ and $\delta$ are chosen so that the observed B/C ratio is reproduced.

\subsection{Reconstruction with the Fermi data}
\begin{figure}[t]
	\centering
	\includegraphics[width=10cm,clip]{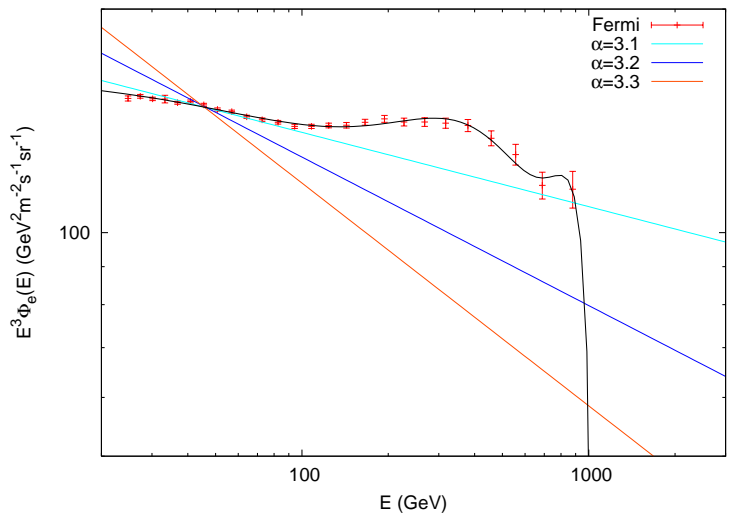} \\
	\caption{The Fermi data~\cite{Fermi} and a fitting curve, together with three different background spectra.}  
	\label{fig:Fermidata}
\end{figure}
As the first example, we consider the Fermi data of the the electron + positron flux~\cite{Fermi}, at energy above 100 GeV. As
the astrophysical background, we take a simple power low $\Phi_e^{\rm bg}\propto E^{-\alpha}$, and vary the index as $\alpha = 3.2\pm 0.1$
to represent the effect of the background uncertainties.\footnote{For instance, for the parametrization of the 
background spectrum in Ref.~\cite{Baltz:1998xv}, based on the simulations of Ref.~\cite{Moskalenko:1997gh}, the high energy
electron + positron spectrum is  
well approximated by a single power $\Phi_{e}^{\rm bg} = \Phi_{e^-}^{\rm bg,prim} + \Phi_{e^-}^{\rm bg,sec} +
\Phi_{e^+}^{\rm bg,sec}\simeq \Phi_{e^-}^{\rm bg,prim} \propto E^{-3.25}$.} 
The normalization is determined by fitting the data for $E< 100$ GeV, assuming that the electron + positron flux is dominated by
the background in this energy range.\footnote{The PAMELA data~\cite{Adriani:2008zr} shows that the positron fraction is less
than ${\cal O}(10)$ \% for $E\lsim 100$ GeV. Assuming that the excess is mainly from the DM decay/annihilation, which generates the same
amount of electron and positron, at least about 80 \% of the total flux is the background in this energy range.}
For illustration, we fit the observed data by a polynomial as $E^3 \Phi_e(E) = \sum_{n=0}^{n_{\rm max}} c_n E^n$ with
$n_{\rm max} = 5$, which is shown in Fig.~\ref{fig:Fermidata} together with three different background spectra 
($\alpha = 3.1$, 3.2, and 3.3).

\begin{figure}[t]
	\centering
	\includegraphics[width=7cm,clip]{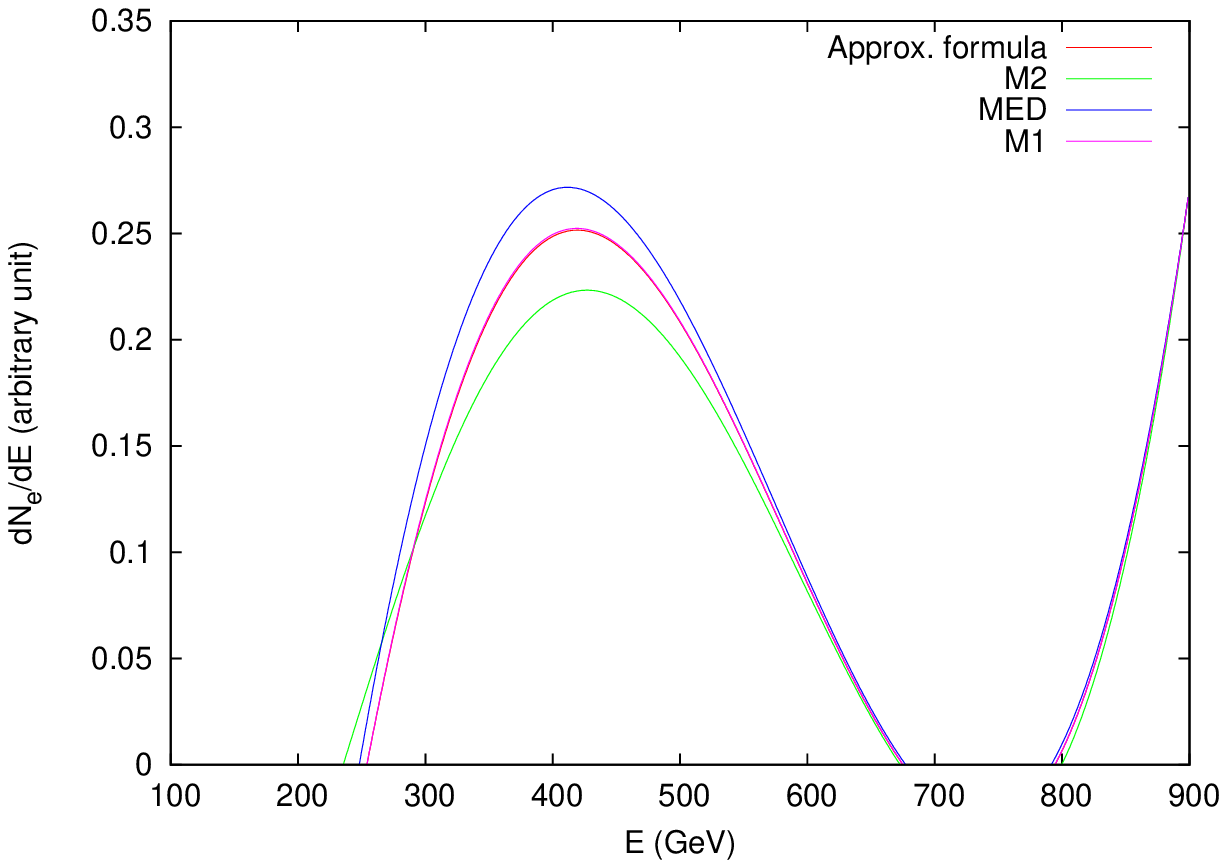}
	\includegraphics[width=7cm,clip]{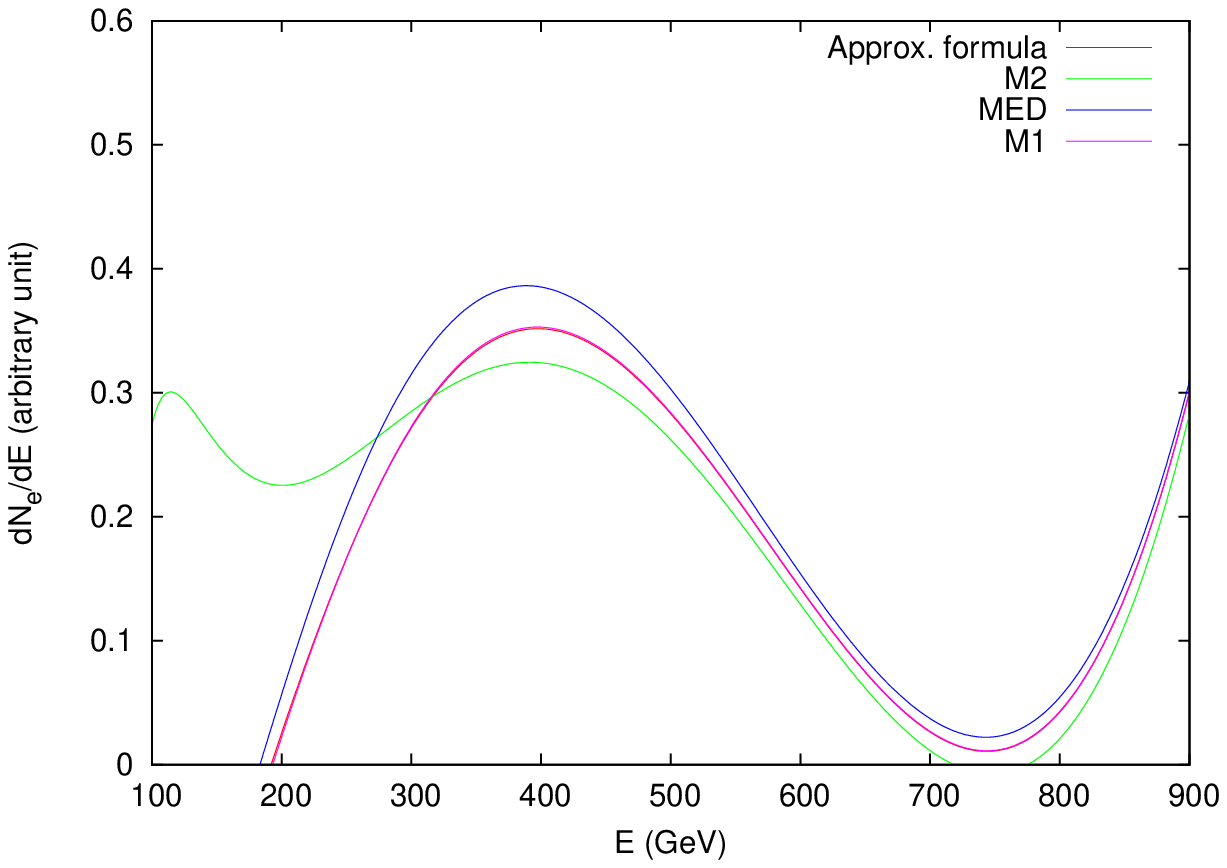}
	\\
	\includegraphics[width=7cm,clip]{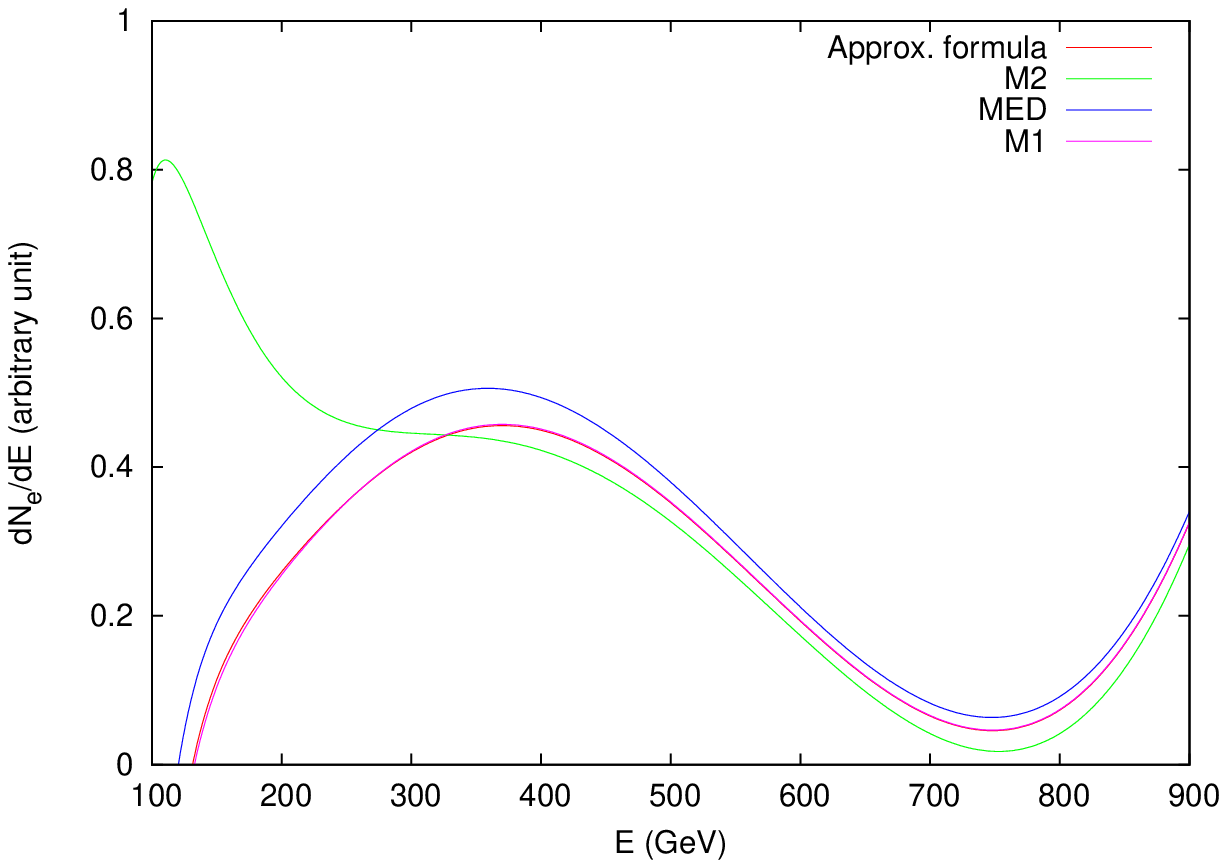}
	\caption{The source spectra reconstructed from the Fermi data in Fig.~\ref{fig:Fermidata}, for different background
	spectrum indices $\alpha = 3.1$ (top left), 3.2 (top right), and 3.3 (bottom). In each case, the results of the analytic
	formula [(\ref{solution_Fourier}) or (\ref{solution_Laplace})] are shown for the decaying DM with the three propagation models.
	The results for the simple approximation formula (\ref{eq:simple}) are also shown, which almost overlap with the M1 model lines in this figure.}
	\label{fig:FermiSource}
\end{figure}
In Fig.~\ref{fig:FermiSource}, the source spectra reconstructed by the analytic formula [(\ref{solution_Fourier}) or (\ref{solution_Laplace})]
are shown for the decaying DM, with the three background spectra and the three propagation models. Here and hereafter,  we assume
the isothermal DM distribution $\rho(|\vec{r}|) = \rho_\odot(r_c^2 + r_\odot^2)/(r_c^2 + |\vec{r}|^2)$
with $r_c=3.5$ kpc. 
We also assume that there is no excess above the energy at which the fitting curve crosses the background.
Interestingly, in all cases there is a clear but broad peak around 400 GeV.
There is another increase of the spectrum for $E\gsim 800$ GeV, which is due to the last two data points (see Fig.~\ref{fig:Fermidata}). 
As can be seen from the figure, the reconstructed source spectrum is almost independent of
the propagation models, except for the energy range $E\lsim 200$ GeV for the M2 model with background index $\alpha \gsim 3.2$.
In Fig.~\ref{fig:FermiSource}, we also show the approximated formula (\ref{eq:simple}).
As discussed in the previous section, the simple approximation formula (\ref{eq:simple}) well reproduces the results of the full formula.
In some cases, the reconstructed source spectrum becomes unphysical (negative) in the energy range of $E\lsim 200$ GeV
and $700\lsim E\lsim800$ GeV.
This is because of the relatively hard spectrum of the observed flux in this energy range. (The simple formula (\ref{eq:simple})
implies that an observed flux much harder than $E^{-2}$ is difficult to obtain from the DM decay/annihilation.)

\begin{figure}[t]
	\centering
	\includegraphics[width=7cm,clip]{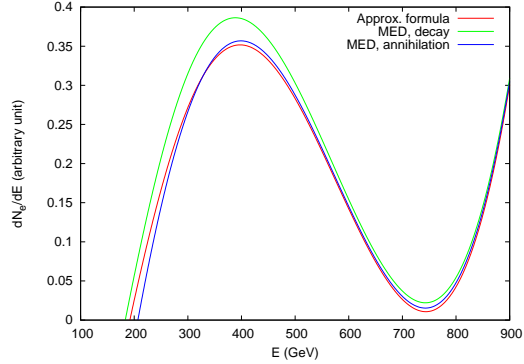}
	\caption{The comparison of the source spectra for the decaying and annihilating DM, reconstructed from the Fermi data in
	Fig.~\ref{fig:Fermidata} with a background spectrum index $\alpha = 3.2$ and MED propagation. The results for the simple
	approximation formula (\ref{eq:simple}) are also shown.}	
\label{fig:FermiComparison}
\end{figure}

In order to see the dependence on the DM distribution (or whether DM is decaying or annihilating), we show in Fig.~\ref{fig:FermiComparison}
the case of annihilating DM compared with the decaying DM, for MED propagation with the background spectrum $\alpha=3.2$.
As expected, the reconstructed spectrum is almost independent of whether the DM is decaying or annihilating.
This is also understood from the approximated formula (\ref{eq:simple}).

\subsection{Reconstruction with the HESS data}
\begin{figure}[t]
	\centering
	\includegraphics[width=10cm,clip]{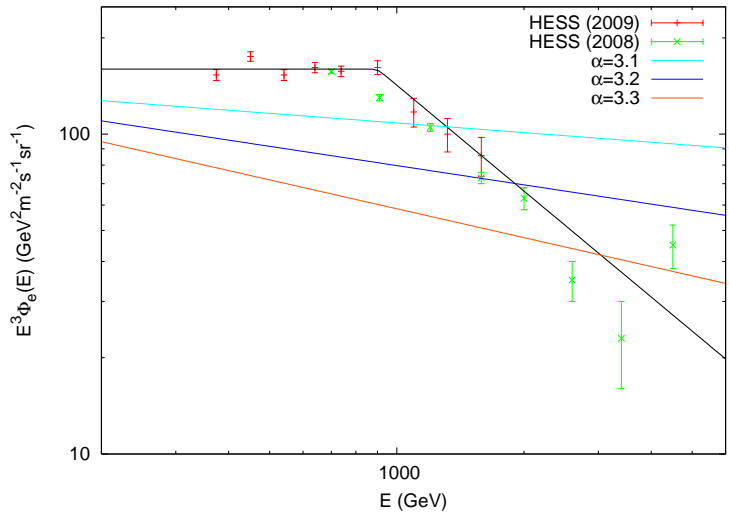} \\
	\caption{The HESS data and a fitting curve~\cite{HESS}, together with the three different background spectra.}  
	\label{fig:HESSdata}
\end{figure}
\begin{figure}[h!]
	\centering
	\includegraphics[width=7cm,clip]{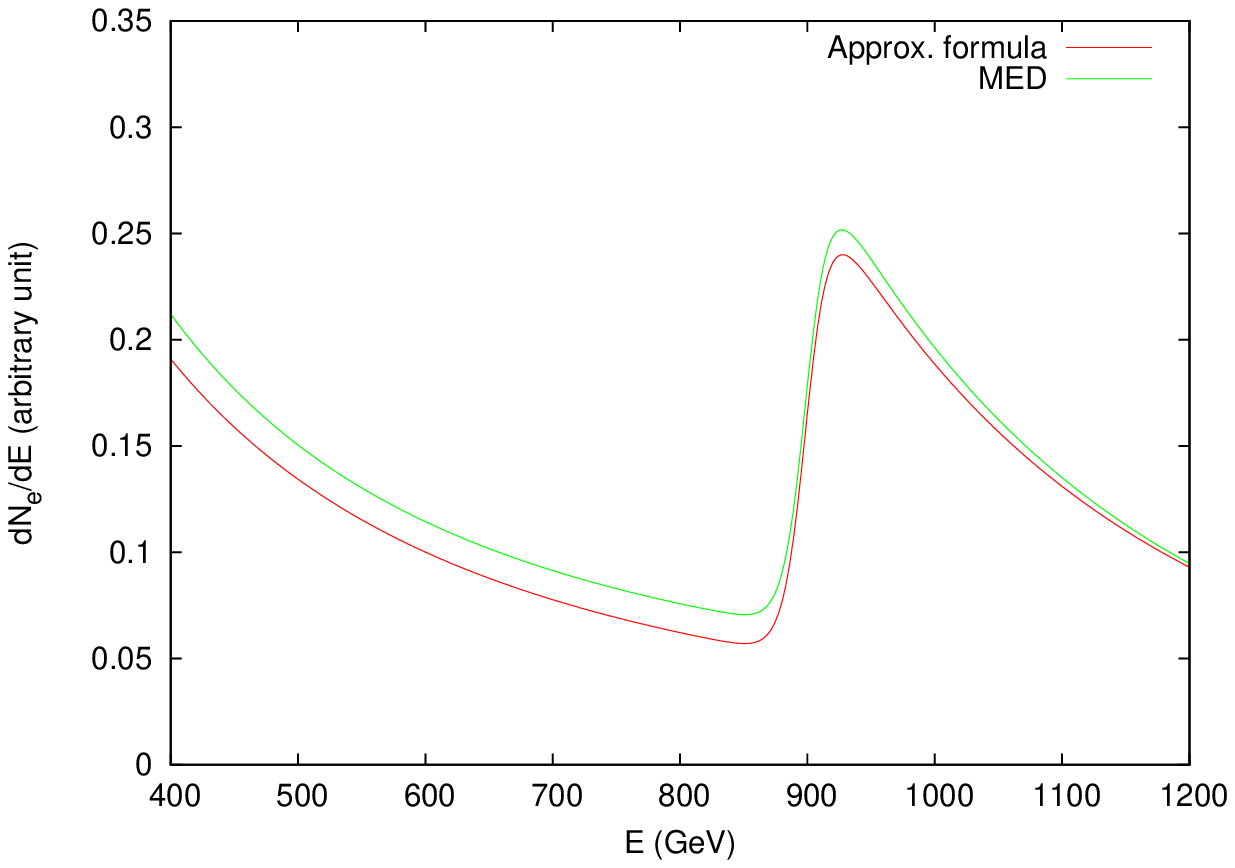}
	\includegraphics[width=7cm,clip]{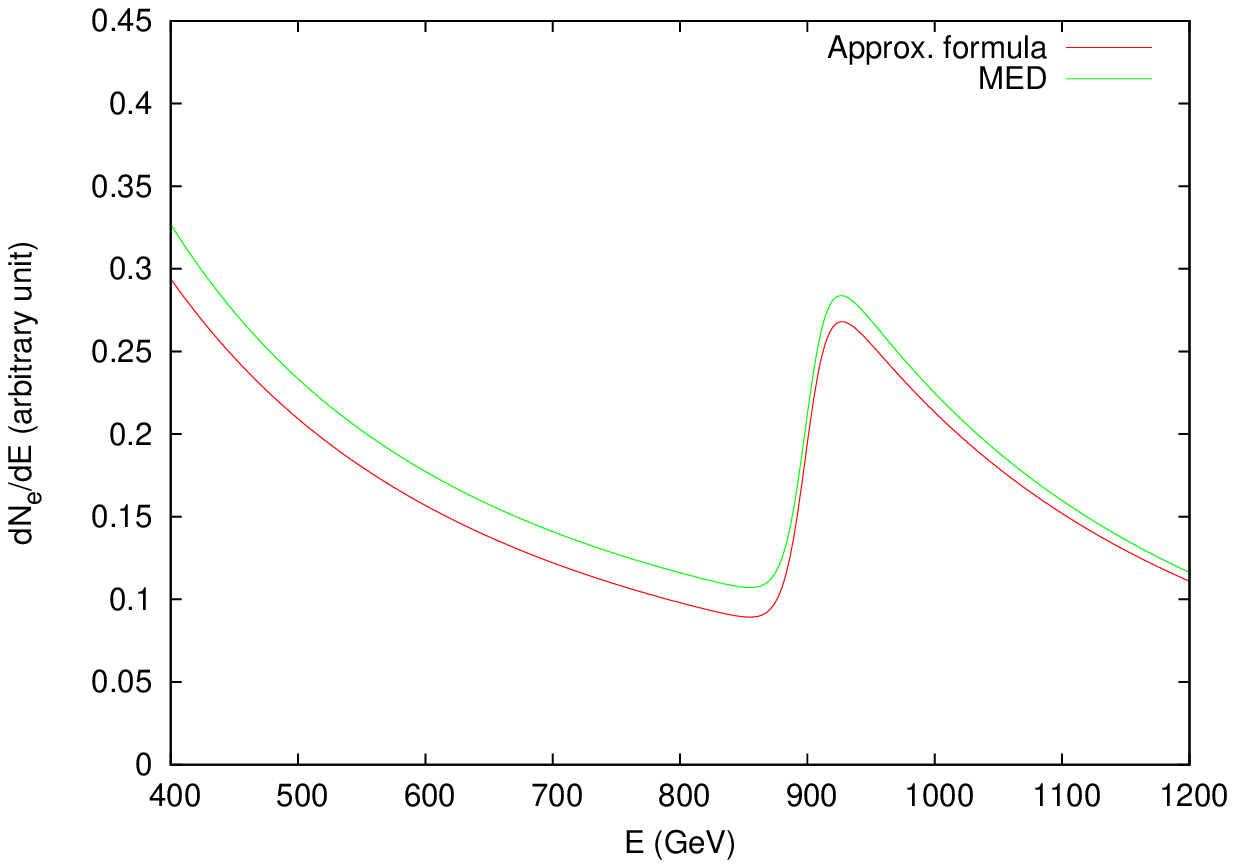}
	\\
	\includegraphics[width=7cm,clip]{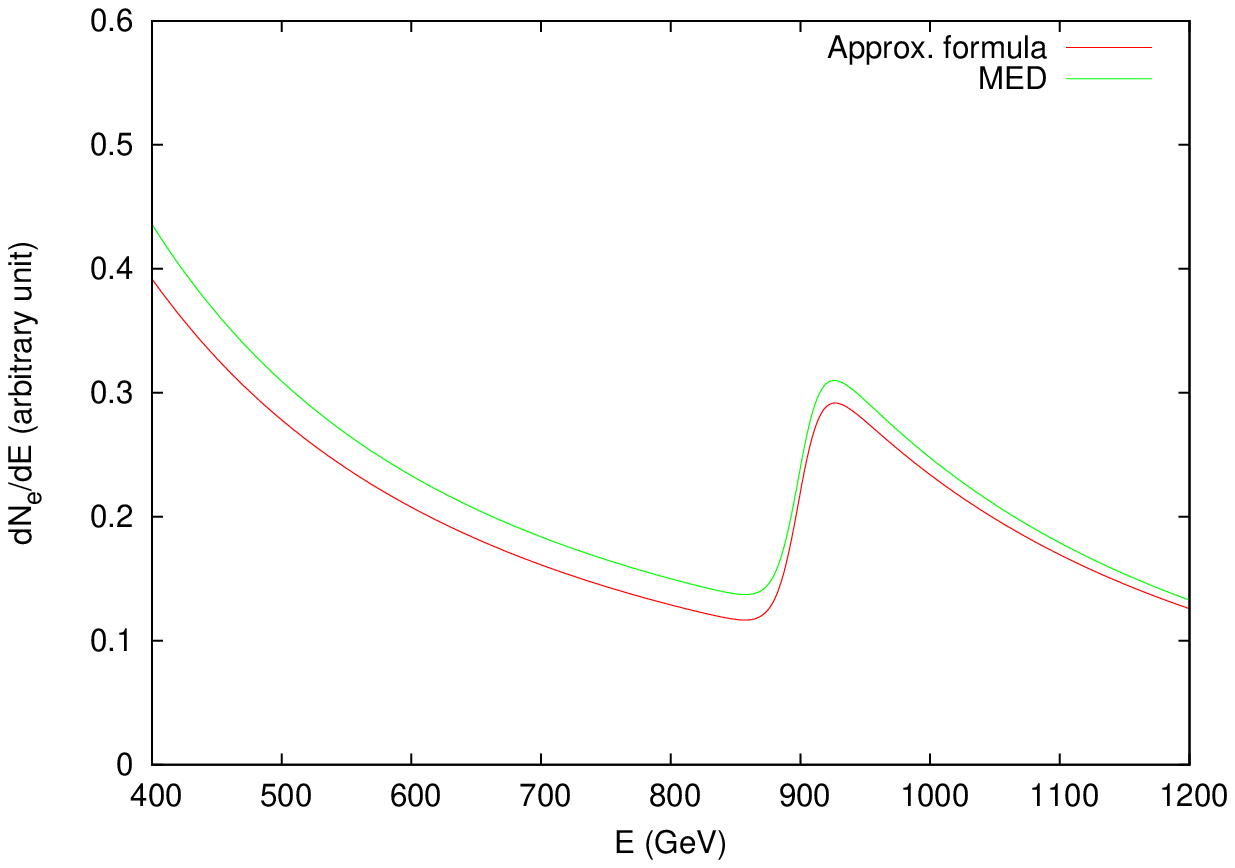}
	\caption{The source spectra reconstructed from the HESS data in Fig.~\ref{fig:HESSdata}, for different background spectrum indices
	$\alpha = 3.1$ (top left), 3.2 (top right), and 3.3 (bottom). In each case, the results of the analytic
	formula [(\ref{solution_Fourier}) or (\ref{solution_Laplace})] are shown for the decaying DM with the MED propagation models,
	together with the simple approximation formula (\ref{eq:simple}).}
	\label{fig:HESS_source}
\end{figure}
Next, we apply the formula to the newly published HESS data~\cite{HESS}, shown in Fig.~\ref{fig:HESSdata}.
As for the fitting function, we adopt the broken power law described in~\cite{HESS}.
We assume the same background spectra $E^{-(3.2\pm 0.1)}$ as before. 
The resultant source spectrum is shown in Fig.~\ref{fig:HESS_source} for decaying DM, isothermal DM profile and MED propagation model. 
One can see that the breaking of the power at $E\sim 1$ TeV results in a peak at that energy in the reconstructed source spectrum.

\subsection{Estimation of the uncertainties}
\begin{figure}[t]
	\centering
	\includegraphics[width=7cm,clip]{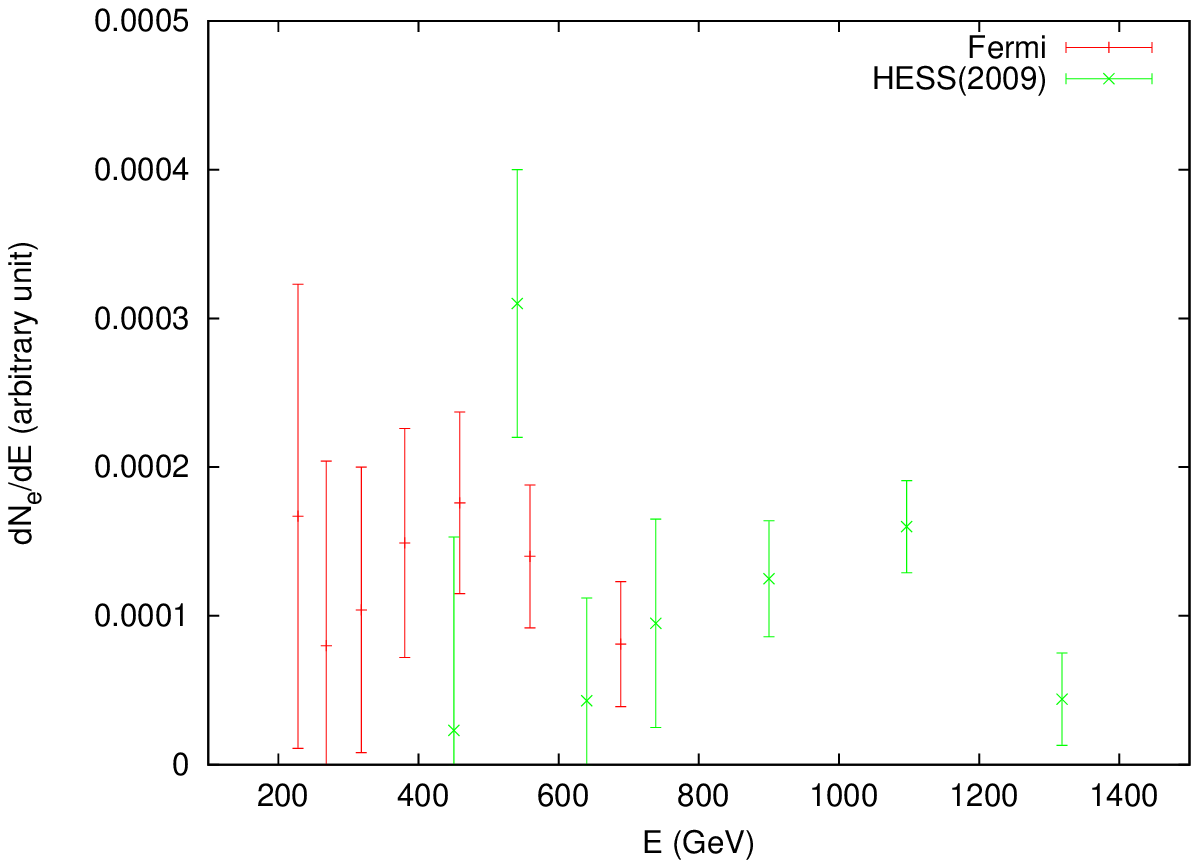} 
	\includegraphics[width=7cm,clip]{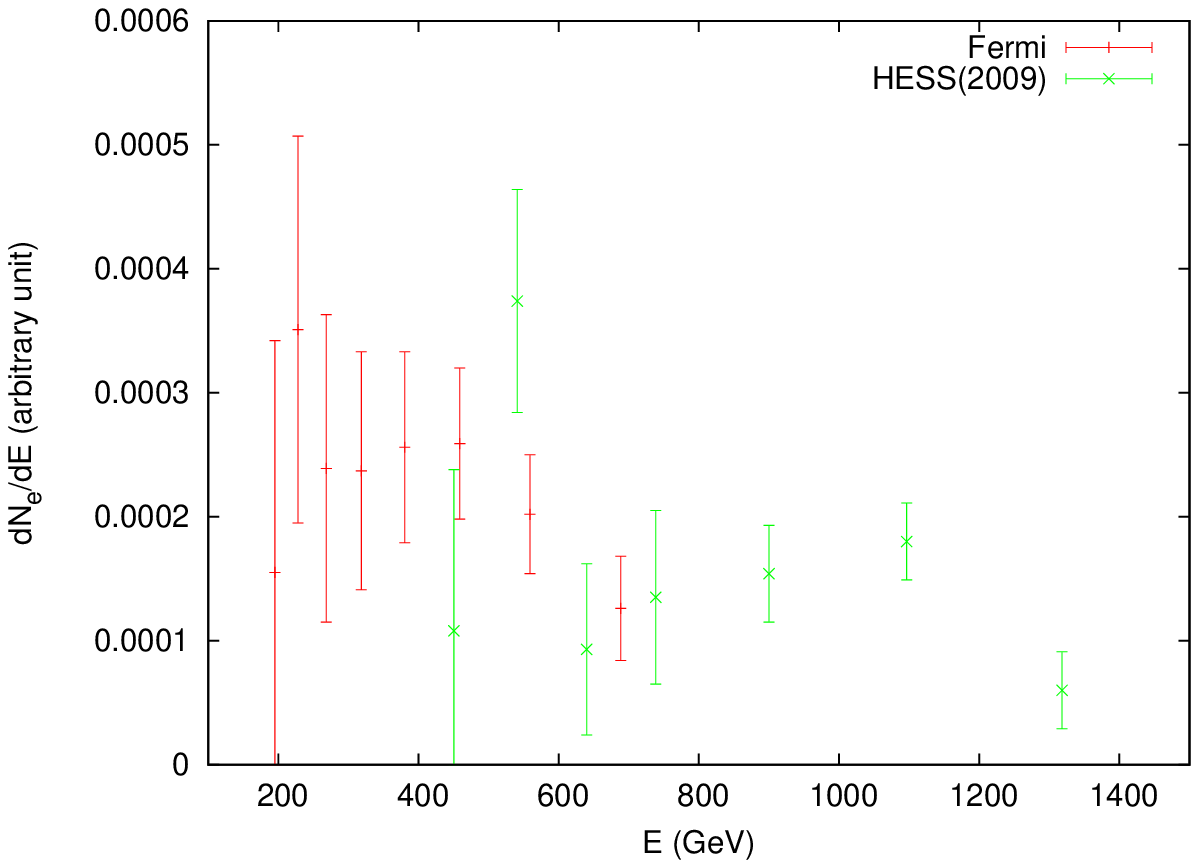} 
	\\
	\includegraphics[width=7cm,clip]{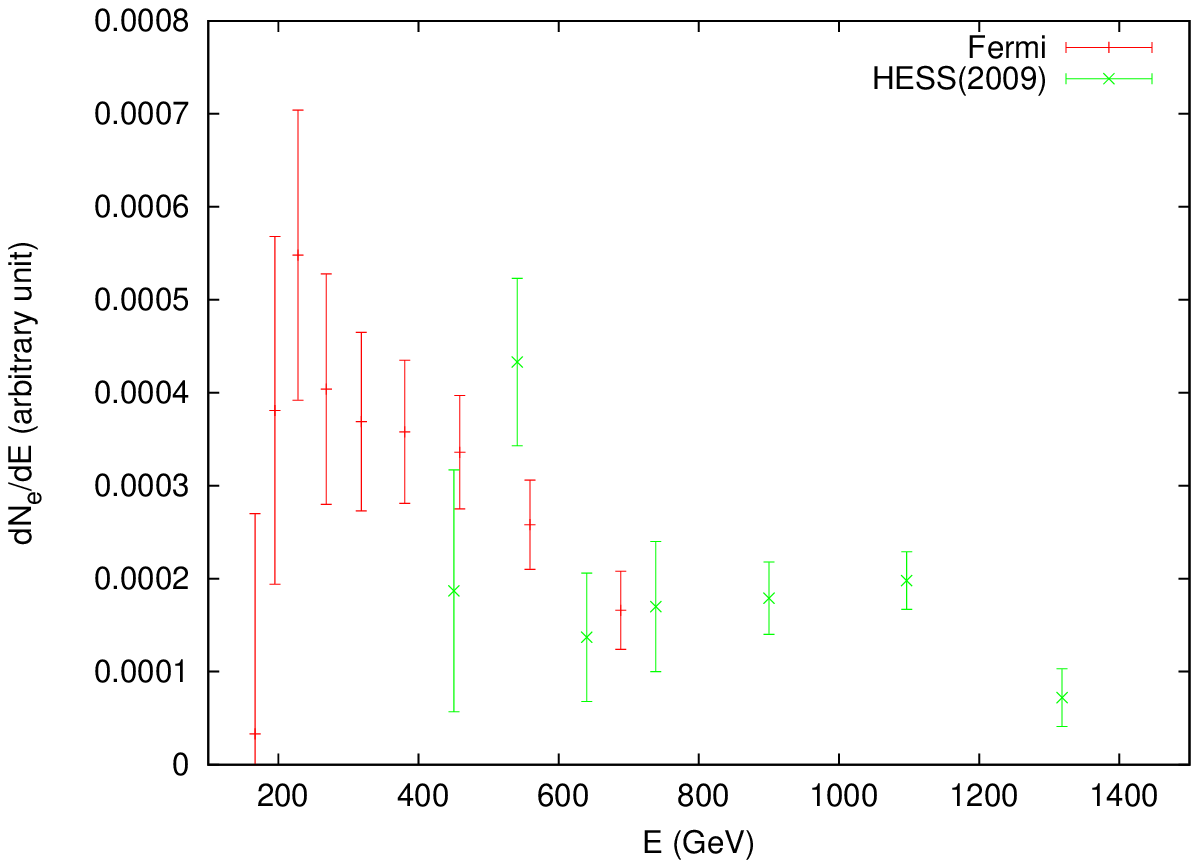}
	\caption{The reconstructed spectrum for the Fermi and HESS data, with errors corresponding to the statistical errors of their
	data~\cite{Fermi,HESS},	with background spectrum indices $\alpha = 3.1$ (top left), 3.2 (top right), and 3.3 (bottom). See text for details.}  
	\label{fig:withError}
\end{figure}
So far, we have neglected the errors in the observed flux. Here, we estimate the effects of the uncertainties in the observed flux.
As we have seen in the previous two subsections, the simple formula (\ref{eq:simple}) is a good approximation in the high energy
range $E\gsim 200$ GeV. Therefore, we use it to estimate the errors of the reconstructed spectrum.
The derivative $dA(E)/dE$ and the corresponding error $\delta[dA(E)/dE]$ at $E=E_i$
are estimated by fitting the sequential three data points at
$E=(E_{i-1}, E_i, E_{i+1})$ with a quadratic function. The resultant source spectrum is shown in Fig.~\ref{fig:withError} for the
Fermi and HESS data, with the three background spectra. Here, we only adopt
the statistical errors of the Fermi and HESS data~\cite{Fermi,HESS}. The broad peaks around 400 GeV and around 1 TeV remain even
after taking into account the errors in the observed flux. 
Although it is difficult to precisely reconstruct the source spectrum because of the limited data as well as the lack of the knowledge
of the astrophysical background, Fig.~\ref{fig:withError} implies that the electrons/positrons from DM have a broad spectrum, ranging
from ${\cal O}(1)$~TeV to ${\cal O}(100)$~GeV. For instance, a direct two-body decay of DM into electron(s), or a three-body decay
into electron(s) with a smooth matrix element, would have harder spectrum and does not fit the reconstructed spectrum well.
The obtained result seems to suggest that the source spectrum has a larger soft component, like the one from cascade decays.

\section{Discussion}
In this letter, we have discussed the possibility of solving the inverse problem of the cosmic-ray electron and positron from the 
dark matter annihilation/decay. Some simple analytic formulae are shown, with which the source spectrum of the electron and positron
can be reconstructed from the observed flux. 
It is shown that the reconstructed source spectrum at an energy $E_{\rm src}$ depends only on
the observed flux above that energy, $E_{\rm obs}\ge E_{\rm src}$. 

We also illustrated our approach by applying the obtained formula to the electron + positron flux above 100 GeV for the just
released Fermi data~\cite{Fermi} and the HESS data~\cite{HESS}. Assuming simple power-law backgrounds, the reconstructed source
spectrum indicates two peaks, a broad one at around 400 GeV and another one at around 1 TeV.

It is shown that the reconstructed spectrum at high energy is almost independent of the propagation models,
and whether it is decaying or annihilating. The effect of the errors in the observed flux is also discussed.
Within the uncertainties, the obtained result implies that the electrons/positrons at the source
have a broad spectrum ranging from ${\cal O}(1)$~TeV
to ${\cal O}(100)$~GeV, with a large soft component, like the one from cascade decays.

It is difficult to precisely reconstruct the source spectrum at the present stage, because of the limited data as well as the lack of
the knowledge of the astrophysical background. Future measurements, such as the PAMELA data at higher energy range, will allow
better understandings. There is also a proposed experiment CALET~\cite{Torii:2006qb}, which can measure the  electron + positron
flux up to 10 TeV with a significant statistics. (cf.~\cite{Chen:2008fx}).
We expect that our approach will be a useful tool in shedding light on the mystery of DM.

\section*{Acknowledgements}

We thank Satoshi Shirai and Tsutomu Yanagida for helpful discussions and comments.
This work was supported by World Premier International Center Initiative
(WPI Program), MEXT, Japan.
The work of EN is supported in part by JSPS Research Fellowships for Young Scientists.

\appendix

\section{Solution to the diffusion equation}
The diffusion equation (\ref{eq:diffusion}) can be solved in the following way (cf. Ref.~\cite{Hisano:2005ec}.)
Using the cylindrical coordinate, $f_e(E,\vec{r})=f_e(E,r,z)$, where $r=\sqrt{x^2+y^2}$ and
\begin{eqnarray}
\nabla^2=\frac{\partial^2}{\partial r^2}+\frac{1}{r}\frac{\partial}{\partial r}
+\frac{\partial^2}{\partial z^2},
\end{eqnarray}
the diffusion equation~(\ref{eq:diffusion}) becomes
\begin{eqnarray}
K(E)\left[\frac{\partial^2}{\partial r^2}+\frac{1}{r}
\frac{\partial}{\partial r}+\frac{\partial^2}{\partial z^2}\right]f_e(E,r,z)
+\frac{\partial}{\partial E}[b(E) f_e(E,r,z)]
+q(r,z)\frac{dN_e(E)}{dE} = 0.
\end{eqnarray}
We expand $f_e$  as
\begin{eqnarray}
f_e(E,r,z)&=&\sum_{n,m=1}^\infty f_{nm}(E) J_0\left(j_n\frac{r}{R}\right)
\sin\Big(\frac{m\pi}{2L}(L-z)\Big),
\\
f_{nm}(E) &=&
\frac{2}{J_1^2(j_n)\pi}\int_0^1d\hat{r}\,\hat{r}
\int_{-\pi}^\pi d\hat{z}\,
J_0(j_n\hat{r})\sin\left(\frac{m}{2}(\pi{-}\hat{z})\right)
f_e\left(E,R\hat{r},\frac{L}{\pi}\hat{z}\right),
\end{eqnarray}
where $J_0$ and $J_1$ are the zeroth and first order Bessel functions of the first kind, 
respectively, and $j_n$ are the successive zeros of $J_0$.
In this expansion, the boundary condition $f_e(E,r,z)=0$ for $r=R$ and $z=\pm L$ is 
automatically satisfied. Conversely, any (sufficiently good) function which satisfies the
boundary condition can be expanded as above. The orthogonal relations are
\begin{eqnarray}
&& \int_0^1dx\,xJ_0(j_n x)J_0(j_{n'}x) = \frac{1}{2}J_1^2(j_n)\delta_{nn'},
\\
&& \int_{-\pi}^\pi dx\,{\rm sin}\left(\frac{m}{2}(\pi{-}x)\right)
{\rm sin}\left(\frac{m'}{2}(\pi{-}x)\right)=\pi\delta_{mm'}\quad(m,m'\in\mathbb{Z}_{>0}).
\end{eqnarray}
Using the differential equation
\begin{eqnarray}
\left[\frac{d^2}{dx^2}+\frac{1}{x}\frac{d}{dx}+j_n^2\right]J_0(j_n x)=0,
\end{eqnarray}
one obtains
\begin{eqnarray}\label{eq_for_fnm}
\left[-d_{nm}K(E)+\frac{\partial b(E)}{\partial E}+b(E)
\frac{\partial}{\partial E}\right]f_{nm}(E)+q_{nm} \frac{dN_e(E)}{dE}=0,
\end{eqnarray}
where $d_{nm}$ and $q_{nm}$ are given by Eqs.~(\ref{eq:dnm}) and (\ref{eq:qnm}).
Imposing the boundary condition for $f_{nm}(E)$ as
\beq
\label{fmn_eq}
f_{nm}(E_{\rm max})=0\quad\text{where}\quad
E_{\rm max}={\rm max}\{E|Q_{nm}(E)\ne0\},
\eeq
the solution to Eq.~(\ref{eq_for_fnm}) is given by
\beq
f_{nm}(E)=\frac{1}{b(E)}\int_E^{E_{\rm max}}dE'\,
\frac{dN_e(E')}{dE'}q_{nm}
{\rm exp}\left[-d_{nm}\left(L(E')-L(E)\right)\right],
\eeq
where $L(E)$ is given by Eq.~(\ref{eq:L}).
This leads to Eq.~(\ref{eq:dN/dE}).

\section{Solution to the inverse problem}\label{inv_sol}
Here we show two ways of reconstructing the source spectrum $dN_e(E)/dE$.

By a change of variables, the equation (\ref{eq:dN/dE}) is rewritten as
\begin{eqnarray}\label{int_eq}
\tilde{A}(x)&=&
-
\int_0^x
dy
\frac{d\tilde{N}_e(y)}{dy}
g(x-y),
\end{eqnarray}
where 
$x=-L(E)$,
$y=-L(E')$,
$\tilde{N}_e(y={-}L(E'))=N_e(E')$, $A(E)=f_e(E)b(E)$, and
$\tilde{A}(x={-}L(E))=A(E)$.
We assumed $L(E_{\rm max})=0$.
Here, $\tilde{A}(x)$ and $g(x)$ are known functions once we fix the propagation model
and have the observational data.
Our inverse problem is now reduced to the problem of solving for the function $d\tilde{N}_e(x)/dx$,
given the functions $\tilde{A}(x)$ and $g(x)$.

The integral equation (\ref{int_eq}) is of the form known as the Volterra's integral equation.
It is known that the solution of the equation exists and is unique.

We see that the right-hand side of Eq.~(\ref{int_eq}) is a convolution of two functions,
so the Fourier transform or the Laplace transform seems to be useful tools.
In the following, we solve the equation in two different ways using
the Fourier and Laplace transforms.

\subsection*{Fourier transform}

Inserting two step functions in the integral, we get
\begin{eqnarray}
\tilde{A}(x)&=&
- 
\int_{-\infty}^{\infty}
dy
\frac{d\tilde{N}_e(y)}{dy}
\theta(y) \
g(x-y)
\theta(x-y).
\end{eqnarray} 

Operating
$\int_{-\infty}^{\infty}
dx
e^{-ikx}
$
on each side of the equation and changing the variable as $z=x-y$ in the right-hand side, we obtain
\begin{eqnarray}
\int_{-\infty}^{\infty}
dx
e^{-ikx}
\tilde{A}(x)
&=&
- 
\int_{-\infty}^{\infty}
dy
e^{-iky}
\frac{d\tilde{N}_e(y)}{dy}
\theta(y) \
\int_{-\infty}^{\infty}
dz
e^{-ikz}
g(z)
\theta(z).
\end{eqnarray} 
After dividing each side by 
$\int_{-\infty}^{\infty}
dz
e^{-ikz}
g(z)
\theta(z)
$
and operate
$
\int_{-\infty}^{\infty}
\frac{dk}{2\pi}
e^{ikx}
$
, we finally get the formula we desire as
\begin{eqnarray}
\frac{d\tilde{N}_e(x)}{dx}
\theta(x)
&=&
-
\int_{-\infty}^{\infty}
\frac{dk}{2\pi}
e^{ikx}
\frac{\int_{-\infty}^{\infty}dw
e^{-ikw}\tilde{A}(w)}{\int_{-\infty}^{\infty}dz
e^{-ikz}g(z)
\theta(z)
},
\end{eqnarray}
or
\begin{eqnarray}
\frac{dN_e(E)}{dE}
\theta\left(-L(E)\right)
&=&
\frac{dL(E)}{dE}
\int_{-\infty}^{\infty}
\frac{dk}{2\pi}
e^{-ikL(E)}
\frac{\int_{-\infty}^{\infty}dw
e^{-ikw}\tilde{A}(w)}{\int_{0}^{\infty}dz
e^{-ikz}g(z)}.
\end{eqnarray}

\subsection*{Laplace transform}

Eq.~(\ref{int_eq}) can be solved in an alternative way. We first differentiate the both side
of the equation with respect to $x$ and then divide by $g(0)$. Then we obtain
\begin{equation}\label{Poisson}
F(x)=\varphi(x)-\int_0^xdy\,K(x-y)\varphi(y),
\end{equation}
where
\begin{align}
\varphi(x)&=\frac{d\tilde{N}_e(x)}{dx}, \\
F(x)&=-\frac{1}{g(0)}\frac{d\tilde{A}(x)}{dx},
\end{align}
and
\begin{equation}
K(x)=-\frac{1}{g(0)}\frac{dg(x)}{dx}.
\end{equation}

The Laplace transform of Eq.~(\ref{Poisson}) is
\begin{equation}
{\cal L}F(\xi)={\cal L}\varphi(\xi)-{\cal L}K(\xi){\cal L}\varphi(\xi),
\end{equation}
where we denote the Laplace transform of a function $f(x)$ by ${\cal L}f(\xi)$:
\begin{equation}
{\cal L}f(\xi)=\int_0^\infty dx\,e^{-\xi x}f(x).
\end{equation}
This equation can be solved as
\begin{equation}
\varphi(x)=F(x)+\int_0^xdy\,\Gamma(x-y)F(y),
\end{equation}
where
\begin{equation}
\Gamma(x)={\cal L}^{-1}\left[\frac{{\cal L}K}{1-{\cal L}K}\right](x)
\end{equation}
and ${\cal L}^{-1}$ denotes the inverse Laplace transform. In the original variable, the solution can be written
as
\begin{equation}
\frac{dN_e(E)}{dE}=-\frac{1}{g(0)}\left[\frac{dA(E)}{dE}+
\frac{dL(E)}{dE}\int_E^{E_{\rm max}}dE'\,\frac{dA(E')}{dE'}\Gamma\left(L(E')-L(E)\right)\right].
\end{equation}


\begin{thebibliography}{99}

\bibitem{DM}
See, for reviews,
  G.~Bertone, D.~Hooper and J.~Silk,
  Phys.\ Rept.\  {\bf 405}, 279 (2005)
  [arXiv:hep-ph/0404175].
  \\ 
  C.~Amsler {\it et al.}  [Particle Data Group],
  Phys.\ Lett.\  B {\bf 667} (2008) 1.
  
    
\bibitem{Adriani:2008zr}
  O.~Adriani {\it et al.}  [PAMELA Collaboration],
  Nature {\bf 458} (2009) 607
  [arXiv:0810.4995 [astro-ph]].

    
  
\bibitem{:2008zzr}
  J.~Chang {\it et al.},
  Nature {\bf 456} (2008) 362.


\bibitem{Torii:2008xu}
  S.~Torii {\it et al.}  [PPB-BETS Collaboration],
  arXiv:0809.0760 [astro-ph].


\bibitem{Fermi}
  Fermi LAT Collaboration,
  arXiv:0905.0025 [astro-ph.HE].
  
\bibitem{HESS}  
  F.~Aharonian {\it et al.}  [H.E.S.S. Collaboration],
  Phys.\ Rev.\ Lett.\  {\bf 101} (2008) 261104
  [arXiv:0811.3894 [astro-ph]];
\\
  F.~Aharonian {\it et al.}  [H.E.S.S. Collaboration],
  arXiv:0905.0105 [astro-ph.HE].
  
  
  
\bibitem{post-Fermi}
  L.~Bergstrom, J.~Edsjo and G.~Zaharijas,
  arXiv:0905.0333 [astro-ph.HE];
\\
  S.~Shirai, F.~Takahashi and T.~T.~Yanagida,
  arXiv:0905.0388 [hep-ph];
\\
  P.~Meade, M.~Papucci, A.~Strumia and T.~Volansky,
  arXiv:0905.0480 [hep-ph];
\\  
  D.~Grasso {\it et al.},
  arXiv:0905.0636 [astro-ph.HE];
\\      
  C.~H.~Chen, C.~Q.~Geng and D.~V.~Zhuridov,
  arXiv:0905.0652 [hep-ph].
\\
  X.~J.~Bi, R.~Brandenberger, P.~Gondolo, T.~Li, Q.~Yuan and X.~Zhang,
  arXiv:0905.1253 [hep-ph].
\\
  K.~Kohri, J.~McDonald and N.~Sahu,
  arXiv:0905.1312 [hep-ph].


  
  
\bibitem{Strong:2007nh}
See for example:
  A.~W.~Strong, I.~V.~Moskalenko and V.~S.~Ptuskin,
  Ann.\ Rev.\ Nucl.\ Part.\ Sci.\  {\bf 57} (2007) 285
  [arXiv:astro-ph/0701517].



\bibitem{Delahaye:2007fr}
  T.~Delahaye, R.~Lineros, F.~Donato, N.~Fornengo and P.~Salati,
  Phys.\ Rev.\  D {\bf 77} (2008) 063527
  [arXiv:0712.2312 [astro-ph]].
  
  
\bibitem{Baltz:1998xv}
  E.~A.~Baltz and J.~Edsjo,
  Phys.\ Rev.\  D {\bf 59} (1999) 023511
  [arXiv:astro-ph/9808243].
  

\bibitem{Moskalenko:1997gh}
  I.~V.~Moskalenko and A.~W.~Strong,
  Astrophys.\ J.\  {\bf 493} (1998) 694
  [arXiv:astro-ph/9710124].
  
  

\bibitem{Hisano:2005ec}
  J.~Hisano, S.~Matsumoto, O.~Saito and M.~Senami,
  Phys.\ Rev.\  D {\bf 73} (2006) 055004
  [arXiv:hep-ph/0511118].
  
\bibitem{Torii:2006qb}
  S.~Torii  [CALET Collaboration],
  Nucl.\ Phys.\ Proc.\ Suppl.\  {\bf 150} (2006) 345;
  J.\ Phys.\ Conf.\ Ser.\ {\bf 120}, (2008) 062020.



\bibitem{Chen:2008fx}
  C.~R.~Chen, K.~Hamaguchi, M.~M.~Nojiri, F.~Takahashi and S.~Torii,
  arXiv:0812.4200 [astro-ph].

  
\end{thebibliography}
\end{document}